\documentclass[12pt]{article}

\usepackage{mathrsfs}
\usepackage[T1]{fontenc}
\usepackage{mathpazo}
\usepackage{setspace}
\usepackage{amsfonts}
\usepackage{amssymb}
\usepackage{amsmath}
\usepackage{epsfig}
\usepackage{latexsym}
\usepackage{color}
\usepackage{graphicx}
\usepackage{nicefrac}
\usepackage[latin1]{inputenc}
\usepackage{pstricks}
\usepackage{slashed}
\usepackage{multirow}

\def\hybrid{\topmargin -20pt    \oddsidemargin 0pt
        \headheight 0pt \headsep 0pt
        \textwidth 6.25in       
        \textheight 9.5in       
        \marginparwidth .875in
        \parskip 5pt plus 1pt   \jot = 1.5ex}

\hybrid

\def\baselinestretch{1.2}

\catcode`\@=11

\def\marginnote#1{}
\newcount\hour
\newcount\minute
\newtoks\amorpm
\hour=\time\divide\hour by60 \minute=\time{\multiply\hour by60
\global\advance\minute by-\hour}
\edef\standardtime{{\ifnum\hour<12 \global\amorpm={am}%
        \else\global\amorpm={pm}\advance\hour by-12 \fi
        \ifnum\hour=0 \hour=12 \fi
        \number\hour:\ifnum\minute<10 0\fi\number\minute\the\amorpm}}
\edef\militarytime{\number\hour:\ifnum\minute<10
0\fi\number\minute}

\def\draftlabel#1{{\@bsphack\if@filesw {\let\thepage\relax
   \xdef\@gtempa{\write\@auxout{\string
      \newlabel{#1}{{\@currentlabel}{\thepage}}}}}\@gtempa
   \if@nobreak \ifvmode\nobreak\fi\fi\fi\@esphack}
        \gdef\@eqnlabel{#1}}
\def\@eqnlabel{}
\def\@vacuum{}
\def\draftmarginnote#1{\marginpar{\raggedright\scriptsize\tt#1}}

\def\draft{\oddsidemargin -.5truein
        \def\@oddfoot{\sl preliminary draft \hfil
        \rm\thepage\hfil\sl\today\quad\militarytime}
        \let\@evenfoot\@oddfoot \overfullrule 3pt
        \let\label=\draftlabel
        \let\marginnote=\draftmarginnote
   \def\@eqnnum{(\theequation)\rlap{\kern\marginparsep\tt\@eqnlabel}%
\global\let\@eqnlabel\@vacuum}  }

\def\preprint{\twocolumn\sloppy\flushbottom\parindent 2em
        \leftmargini 2em\leftmarginv .5em\leftmarginvi .5em
        \oddsidemargin -.5in    \evensidemargin -.5in
        \columnsep .4in \footheight 0pt
        \textwidth 10.in        \topmargin  -.4in
        \headheight 12pt \topskip .4in
        \textheight 6.9in \footskip 0pt
        \def\@oddhead{\thepage\hfil\addtocounter{page}{1}\thepage}
        \let\@evenhead\@oddhead \def\@oddfoot{} \def\@evenfoot{} }

\def\numberbysection{\@addtoreset{equation}{section}
        \def\theequation{\thesection.\arabic{equation}}}

\def\underline#1{\relax\ifmmode\@@underline#1\else
        $\@@underline{\hbox{#1}}$\relax\fi}

\def\titlepage{\@restonecolfalse\if@twocolumn\@restonecoltrue\onecolumn
     \else \newpage \fi \thispagestyle{empty}\c@page\z@
        \def\thefootnote{\fnsymbol{footnote}} }

\def\endtitlepage{\if@restonecol\twocolumn \else \newpage \fi
        \def\thefootnote{\arabic{footnote}}
        \setcounter{footnote}{0}}  

\catcode`@=12 \relax

\def\figcap{\section*{Figure Captions\markboth
        {FIGURECAPTIONS}{FIGURECAPTIONS}}\list
        {Figure \arabic{enumi}:\hfill}{\settowidth\labelwidth{Figure
999:}
        \leftmargin\labelwidth
        \advance\leftmargin\labelsep\usecounter{enumi}}}
 \relax
\def\tablecap{\section*{Table Captions\markboth
        {TABLECAPTIONS}{TABLECAPTIONS}}\list
        {Table \arabic{enumi}:\hfill}{\settowidth\labelwidth{Table
999:}
        \leftmargin\labelwidth
        \advance\leftmargin\labelsep\usecounter{enumi}}}
 \relax
\def\reflist{\section*{References\markboth
        {REFLIST}{REFLIST}}\list
        {[\arabic{enumi}]\hfill}{\settowidth\labelwidth{[999]}
        \leftmargin\labelwidth
        \advance\leftmargin\labelsep\usecounter{enumi}}}
 \relax
\makeatletter
\newcounter{pubctr}
\def\publist{\@ifnextchar[{\@publist}{\@@publist}}
\def\@publist[#1]{\list
        {[\arabic{pubctr}]\hfill}{\settowidth\labelwidth{[999]}
        \leftmargin\labelwidth
        \advance\leftmargin\labelsep
        \@nmbrlisttrue\def\@listctr{pubctr}
        \setcounter{pubctr}{#1}\addtocounter{pubctr}{-1}}}
\def\@@publist{\list
        {[\arabic{pubctr}]\hfill}{\settowidth\labelwidth{[999]}
        \leftmargin\labelwidth
        \advance\leftmargin\labelsep
        \@nmbrlisttrue\def\@listctr{pubctr}}}
 \relax
\makeatother
\newskip\humongous \humongous=0pt plus 1000pt minus 1000pt

\newif\ifdtup

\relax

\def\be{\begin{equation}}
\def\ee{\end{equation}}
\def\ba{\begin{eqnarray}}
\def\ea{\end{eqnarray}}

\def\no{\noindent}

\def\IR{\relax{\rm I\kern-.18em R}}

\def\IR{\relax{\rm I\kern-.18em R}}
\def\inv{^{\raise.15ex\hbox{${\scriptscriptstyle -}$}\kern-.05em 1}}

\begin{document}

\renewcommand{\theequation}{\thesection.\arabic{equation}}

\newcommand{\beq}{\begin{equation}}
\newcommand{\eeq}[1]{\label{#1}\end{equation}}
\newcommand{\ber}{\begin{eqnarray}}
\newcommand{\eer}[1]{\label{#1}\end{eqnarray}}
\newcommand{\eqn}[1]{(\ref{#1})}
\begin{titlepage}
\begin{center}

\vskip -.1 cm
\hfill June 2010\\

\vskip .6in

{\large \bf Homogeneous vacua of (generalized) new massive gravity}

\vskip 0.6in

{\bf Ioannis Bakas} and {\bf Christos Sourdis}\footnote{Present address:
Technological Education Institution, 34100 Chalkida, Greece.} \vskip 0.2in
{\em Department of Physics, University of Patras \\
GR-26500 Patras, Greece\\
\vskip 0.2in
\footnotesize{\tt bakas@ajax.physics.upatras.gr, sourdis@on.gr}}\\

\end{center}

\vskip 0.8in

\centerline{\bf Abstract} \no
We obtain all homogeneous solutions of new massive gravity models on
$S^3$ and $AdS_3$ by extending previously known results for the cosmological
topologically massive theory of gravity in three dimensions. In all cases,
apart from the maximally symmetric vacua, there are axially symmetric (i.e.,
bi-axially squashed) as well as totally anisotropic (i.e., tri-axially
squashed) metrics of special algebraic type. Transitions among the vacua
are modeled by instanton solutions of $3+1$ Ho\v{r}ava--Lifshitz gravity
with anisotropic scaling parameter $z=4$.

\vfill
\end{titlepage}
\eject

\def\baselinestretch{1.2}
\baselineskip 16 pt \noindent
\section{Introduction}
\setcounter{equation}{0}

Over the years, there has been considerable interest in toy models of
gravitational physics at the classical and quantum levels by focusing, in
particular, to theories in $2+1$ space-time dimensions (for a modern overview
of the subject see, for instance, \cite{carlip}, and references therein).
Einstein gravity (with or without cosmological constant) has no propagating
degrees of freedom in three dimensions, and, as such, it appears to be of limited
interest at first sight. Nevertheless, it provides a soluble model that has
been studied extensively as topological (Chern--Simons) field theory using a
natural reformulation in terms of the spin connection, \cite{town, witten1}.
Several interesting questions have been addressed in this context, including
the possible resolution of classical singularities and topology changing amplitudes in
the quantum theory, \cite{witten2}. Localized matter sources were also included
and found to affect the geometry globally rather than locally, thus leading
to conical singularities in space-time, \cite{jackiw}. Furthermore, the presence of
a cosmological constant allowed the construction of $AdS_3$ black-hole solutions,
\cite{Banados:1992wn}, and also led to important developments in connection with
two-dimensional conformal symmetries, \cite{brown}, as predecessor of AdS/CFT
correspondence. The interest in the subject was revived recently, while searching
for conformal field theories dual to pure three-dimensional gravity with
negative cosmological constant, \cite{witten3}.

Massive generalizations of three-dimensional gravity provide an interesting
twist as they allow propagating degrees of freedom in space-time. Topologically
massive gravity is the prime example obtained by adding a gravitational
Chern--Simons term to the usual Einstein--Hilbert action in the spirit of
topologically massive gauge theories, \cite{DJT, DJT2}. The model was
extended by the addition of a cosmological constant term to cosmological
topologically massive gravity and it was further generalized to three-dimensional
supergravity, \cite{Deser82}. The gravitational Chern--Simons term is odd under
parity and as a result the theory exhibits a single massive propagating degree
of freedom of a given helicity, whereas the other helicity mode remains
massless. Topologically massive gravity appears to be a renormalizable quantum
field, \cite{stan1} (but see also \cite{oda1} for a more recent discussion), which
makes it a valuable model. Various solutions have been obtained and studied over
the years (see, for instance, \cite{Chow:2009km}, and references therein) including
$AdS_3$ black-holes. The cosmological variant of the theory was also investigated
in detail at the chiral point, \cite{andy}, and in the context of AdS/CFT
correspondence, \cite{skende}, leading to surprisingly rich mathematical structures
that are still under investigation and revived interest in the model.

Another massive generalization of three-dimensional gravity was proposed recently
by adding a specific quadratic curvature term to the Einstein--Hilbert action,
\cite{BHT, Bergshoeff:2009aq}. This term (to be discussed later in detail) was designed to
yield upon linearization the Pauli--Fierz action for a massive propagating graviton, and
the resulting theory became known as new massive gravity. This model also appears to be a unitary
renormalizable quantum field theory in three dimensions, \cite{Nakasone:2009bn, Oda:2009ys},
but unlike topologically massive gravity, the new theory preserves parity, and, as a result,
the gravitons acquire the same mass for both helicity states. Models of this type with
quadratic curvature terms are known to be renormalizable in four
space-time dimensions, \cite{stelle}, which, in turn, imply power counting
super-renormalizability of the corresponding three-dimensional theory. Adding a cosmological
constant term yields the cosmological new massive gravity. Further generalization
is provided by combining the effect of the gravitational Chern--Simons and the
quadratic curvature terms to the Einstein--Hilbert action (with or without
cosmological constant), thus leading to the so called generalized massive gravity
model that encompasses all previously known theories of three-dimensional gravity
in a unified framework. The resulting theory exhibits by construction propagating
degrees of freedom, but with different masses for the two helicity states of the
graviton, which reduce consistently to the graviton modes of the topologically massive
and the new massive theories of gravity in different corners of the space of coupling
constants, \cite{BHT, Bergshoeff:2009aq}. Generalization to three-dimensional
supergravity was subsequently considered, \cite{Andringa:2009yc}, and the chiral
point of generalized massive gravity in $AdS_3$ was investigated, \cite{liusun}.
Also, various solutions of new massive gravity, including $AdS_3$ black-holes,
have already been constructed in the literature, \cite{Clement:2009gq, Oliva:2009ip},
and some aspects of the AdS/CFT correspondence turned out to be on par with the
holographic studies of topologically massive gravity, \cite{daniel}.

The subject of three-dimensional massive gravity provides an active area of research
where new developments or applications are mostly welcome.
The landscape of vacua is quite rich and it has not been explored in all generality.
Various special classes of solutions already exist in the literature -- we do not
intend to list them all -- and some general methods have been proposed for their
explicit construction, \cite{Gurses:2010sm}. Certainly,
solutions of new massive gravity, and its generalizations, are less studied compared
to topologically massive gravity, as they involve certain fourth-order equations in
three-dimensional geometry. Most notably, what is still lacking is the construction of
homogeneous solutions which generalize the maximally symmetric vacuum to anisotropic
model geometries.
The main purpose of the present work is to investigate the class of such locally
homogeneous vacua of generalized massive gravity (with or without cosmological constant)
by focusing, in particular, to Bianchi IX metrics on $S^3$ when the theory is defined in the
Euclidean regime. In this case, we will be able to solve the field equations in all
generality and obtain configurations with different degree of anisotropy. As
will be seen later, these solutions reduce to the homogeneous solutions of topologically
massive gravity in the appropriate corner of the space of coupling constants, which have
been known for a long time, \cite{vuorio, Nutku:1989qi, Ortiz:1989vc} (but see also
\cite{Chow:2009km}). Likewise, by analytic continuation, we will obtain all
homogeneous metrics on $AdS_3$ in the Lorentzian version of the theory.

The interest in the solutions of three-dimensional massive gravity also stems
from the fact that they provide static solutions of $3+1$ Ho\v{r}ava--Lifshitz
gravity. Recall that the non-relativistic theory of gravity proposed recently by
Ho\v{r}ava, \cite{horava}, involves the Euclidean action of three-dimensional gravity
as superpotential functional (assuming detailed balance). Using the generalized
three-dimensional massive gravity model, one obtains a super-renormalizable version of
Ho\v{r}ava--Lifshitz gravity in $3+1$ space-time dimensions with anisotropic scaling
$z=4$  (see also \cite{cai}), whereas restriction to topologically massive gravity
yields a non-relativistic theory with anisotropic scaling parameter $z=3$.
Following \cite{bakas}, it is possible to consider gravitational instantons
of Ho\v{r}ava--Lifshitz theory, which are defined as eternal solutions of the gradient flow
equations following from the action of three-dimensional massive gravity. Then,
instantons with $SU(2)$ isometry, which are the simplest to consider, are determined by
first reducing the flow equations to three--dimensional geometries of Bianchi
type IX and then classifying all eternal solutions that interpolate between the fixed points.
The homogeneous vacua of generalized three-dimensional massive gravity are not only
solutions of Ho\v{r}ava--Lifshitz gravity, but they also offer the end points (as fixed
points of the flow lines) to support such instanton solutions. Therefore, the results we
report here can also be used as starting point to extend the methods of our previous
work \cite{bakas} to the construction of gravitational instantons in non-relativistic
theories with higher anisotropic scaling parameter.

The material of this paper is organized as follows: In section 2, we briefly review the
theories of three-dimensional massive gravity, with increasing degree of complexity, and
their field equations, setting up the notation and the framework of our study. In section 3,
we introduce the Bianchi IX ansatz by considering metrics on $S^3$ with $SU(2)$ isometry
group and solve the field equations of Euclidean generalized massive gravity. The solutions
are classified and tabulated according to the geometric characteristics of the metrics and
the limiting cases of new massive gravity and topologically massive gravity are discussed
separately. These results constitute the main body of our work.
In section 4, the solutions are extended to the Lorentzian regime by analytic
continuation of the metrics and obtain all homogeneous solutions of generalized massive
gravity on $AdS_3$. The resulting configurations are also characterized algebraically using
the general classification schemes of Petrov and Serge. In section 5, the results are taken
in the context of Ho\v{r}ava--Lifshitz gravity and we outline the construction and
classification of the corresponding $SU(2)$ gravitational instantons when the anisotropic
scaling parameter of the theory is $z=4$. Finally, in section 6, we present the conclusions
and some directions for future work.

\section{Massive gravity in three dimensions}
\setcounter{equation}{0}

The models of three-dimensional massive gravity are based on certain higher-order
extensions of pure Einstein gravity. We first consider the Einstein--Hilbert action,
\be
S_{\rm EH} = {1 \over \kappa_{\rm w}^2} \int d^3 x \, \sqrt{|g|} \left( R-2\Lambda \right) ,
\ee
including also the effect of a cosmological constant $\Lambda$, which can assume any
value. The three-dimensional gravitational coupling $\kappa_{\rm w}$ will be normalized
to 1 for convenience, but it can be easily reinstated by rescaling the other couplings.
We have three massive theories of gravity that are presented in increasing order of
complexity.

{\bf Topologically massive gravity}: It is defined by adding the gravitational
Chern--Simons term to the Einstein--Hilbert action, following \cite{DJT,DJT2},
\be
S_{\rm{TMG}}= S_{\rm EH} + {1 \over \omega} S_{\rm CS} ~,
\ee
where
\be
S_{\rm CS} = \frac{1}{2} \int d^3 x \, \sqrt{|g|} ~ \varepsilon^{\lambda\mu\nu}
\Gamma{^\rho}_{\lambda\sigma} \left( \partial_\mu\Gamma{^\sigma}_{\rho\nu}
+\frac{2}{3} \Gamma{^\sigma}_{\mu\tau} \Gamma{^\tau}_{\nu\rho}\right)
\ee
is written in terms of the usual Levi--Civita connection of the space-time metric $g$.
Here, $\varepsilon^{\mu\nu\rho}$ is the fully anti-symmetric symbol in three
dimensions with $\varepsilon^{123}=1$. Clearly, the Chern--Simons term flips sign under
orientation reversing transformations and the theory is not invariant under parity.

The classical equations of motion are obtained by varying the action with respect
to the metric and they read as
\be
R_{\mu \nu} - \frac{1}{2} R g_{\mu \nu}
+ \Lambda g_{\mu \nu} + {1 \over \omega} C_{\mu \nu} = 0 \, ,
\label{TMGe}
\ee
where $C_{\mu \nu}$ is the Cotton tensor of the metric $g$, which is defined as follows
\be
C_{\mu \nu} = {\varepsilon{_\mu}{^{\rho\sigma}} \over \sqrt{|g|}}
\nabla_\rho (R_{\nu \sigma} - \frac{1}{4} R g_{\nu \sigma})
\ee
and it is a traceless and covariantly conserved symmetric tensor.
Taking the trace of equation \eqn{TMGe} yields $R = 6 \Lambda$ for the classical
solutions, whereas the remaining equations of motion can be cast in the form
\be
R_{\mu \nu} - \frac{1}{3} R g_{\mu \nu} + {1 \over \omega} C_{\mu \nu} = 0
\label{TMGe2}
\ee
that is most appropriate for the algebraic (Petrov--Segre) characterization of the
corresponding solutions, as will be seen later.

Clearly, the maximally symmetric Einstein metrics which are conformally flat are common
solutions of topologically massive gravity with ordinary Einstein theory.

{\bf New massive gravity}: It is defined by adding a very special quadratic curvature
term to the Einstein--Hilbert action, following \cite{BHT, Bergshoeff:2009aq},
\be
S_{\rm{NMG}}= S_{\rm EH} - {1 \over m^2} S_{\rm BHT} ~,
\ee
where the new term (assuming signature $-++$ in the Lorentzian version of the theory rather than $+--$ used in the original works)
\be
S_{\rm{BHT}}= \int d^3 x \, \sqrt{|g|} \left( R_{\kappa\lambda}R^{\kappa\lambda}-
\frac{3}{8}R^2\right)
\label{BHT}
\ee
is denoted by the initials of its inventors. Unlike topologically massive gravity,
this theory preserves parity.

The corresponding classical equations of motion are certain fourth-order equations of the
form
\be
R_{\mu \nu} - \frac{1}{2} R g_{\mu \nu}
+ \Lambda g_{\mu \nu} - \frac{1}{2 m^2}K_{\mu\nu} = 0 \, ,
\label{CNMGe}
\ee
where $K_{\mu\nu}$ is the symmetric and covariantly conserved tensor
\be
K_{\mu\nu} =  2\nabla^2 R_{\mu\nu} - \frac{1}{2}\nabla_{\mu}\nabla_{\nu}R +
\frac{9}{2}RR_{\mu\nu} -8{R_{\mu}}^\kappa R_{\nu\kappa} -
g_{\mu\nu}\left(\frac{1}{2}\nabla^2 R-3 R_{\kappa\lambda}R^{\kappa\lambda}+
\frac{13}{8}R^2\right) .
\label{KK}
\ee
The special feature is that its trace coincides with the Lagrangian density of
$S_{\rm{BHT}}$,
\be
K \equiv g^{\mu\nu} K_{\mu\nu} = R_{\kappa\lambda}R^{\kappa\lambda}-\frac{3}{8}R^2\,,
\label{Ktr}
\ee
which singles out the special value $q = -3/8$ among the more general combination
of quadratic curvature terms $R_{\kappa\lambda}R^{\kappa\lambda} + qR^2$ in the action.

As before, taking the trace of equation \eqn{CNMGe} yields $K + m^2(R-6\Lambda)=0$ for the
classical solutions, whereas the remaining equations of motion can be written as a sum of
two traceless tensors, namely,
\be
R_{\mu \nu} - \frac{1}{3} R g_{\mu \nu} - \frac{1}{2m^2}(K_{\mu \nu} -
\frac{1}{3} K g_{\mu \nu}) = 0 \, ,
\ee
which is also a useful form for the algebraic classification of the corresponding metrics.

{\bf Generalized massive gravity}: It is obtained by combining the Einstein--Hilbert action
with both higher order terms in the form
\be
S_{\rm{GMG}}= S_{\rm EH} + {1 \over \omega} S_{\rm CS} - {1 \over m^2} S_{\rm BHT} ~,
\ee
thus, providing the most general massive theory of gravity up to four derivative
terms\footnote{An extension of new massive gravity to all orders in the curvature was
recently proposed via a gravitational Born--Infeld action, \cite{tekin1}. Likewise, a
Born--Infeld extension of $(3+1)$-dimensional Ho\v{r}ava--Lifshitz gravity was proposed
to account for arbitrary anisotropy scaling parameter $z$ in the curvature expansion,
\cite{tekin2}. We will not investigate such generalized massive theories of gravity here,
since we are only limited to models with up to four derivative terms, but, clearly, it
is interesting to inquire about locally homogeneous (and other) solutions of the
gravitational Born--Infeld theory.}. Clearly, it is not invariant under parity, in
general, and reduces to all simpler massive gravity models in the appropriate corners
of the space of coupling constants.

In this general case, the classical equations of motion take the following form
\be
R_{\mu \nu} - \frac{1}{2} R g_{\mu \nu}
+ \Lambda g_{\mu \nu} - \frac{1}{2 m^2}K_{\mu\nu} + {1 \over \omega} C_{\mu \nu} = 0
\label{CGMGe}
\ee
and include a mixture of second, third and fourth order derivative terms, making them
more intricate to study. The trace of equation \eqn{CGMGe} yields $K + m^2(R-6\Lambda)=0$,
which is the same condition as for new massive gravity, whereas the remaining components
can be organized as the sum of three traceless tensors,
\be
R_{\mu \nu} - \frac{1}{3} R g_{\mu \nu} - \frac{1}{2m^2}(K_{\mu \nu} -
\frac{1}{3} K g_{\mu \nu}) + {1 \over \omega} C_{\mu \nu} = 0 \, .
\label{CGMGe2}
\ee

We conclude the general presentation of these models by considering three special
limiting cases that may arise in the space of coupling constants. Pure second order
Einstein gravity arises in the limit $\kappa_{\rm w} \rightarrow 0$ with Einstein metrics
as vacua. Pure third order Cotton theory is conformal gravity that arises in the
limit $\omega \rightarrow 0$ and has conformally flat metrics as vacua; it admits
an alternative Chern--Simons gauge field interpretation based on the conformal group
in three dimensions, \cite{horne}. Finally, a special fourth order
theory arises in the limit $m \rightarrow 0$, which has already been studied in the
literature as ghost-free model of gravitation, \cite{standes}. As will be seen later,
these three special limiting cases admit some very simple homogeneous solutions that
form the base for other vacua. For more general values of the couplings, the vacua
arise by balancing three competing terms of different order and their form can be
rather complex.

\section{Homogeneous solutions on $S^3$}
\setcounter{equation}{0}

In this section we construct and classify all locally homogeneous solutions of the
Euclidean generalized massive gravity on $S^3$. The homogeneous vacua of
new massive gravity, as well as topologically massive gravity (which are already known
in the literature) will follow from the general expressions as limiting cases.

\subsection{Bianchi IX model geometries}

First, we present some background material for homogeneous geometries with isometry group
$SU(2)$ based on Bianchi classification (see, for instance, \cite{ExactSol}). We will
also compute the components of the curvature tensors $R_{ij}$, $C_{ij}$ and $K_{ij}$ for
this class of models. Such
geometries provide consistent reduction of the field equations to an algebraic system of
equations of three variables that turns out to be exactly soluble.

The line element of locally homogeneous geometries on $S^3$ takes the following form
\begin{equation}
\label{3d}
\mathrm{d}s^2 = \gamma_1 ~ \sigma_1^2 + \gamma_2 ~ \sigma_2^2 + \gamma_3 ~ \sigma_3^2  ~,
\end{equation}
using the left-invariant Maurer--Cartan one-forms of $SU(2)$, $\sigma_i$, which
satisfy the relations
\begin{equation}
\mathrm{d}\sigma_i + \frac{1}{2} {\varepsilon_i}^{jk}\sigma_j \wedge \sigma_k=0 ~.
\end{equation}
More explicitly, in terms of Euler angles ranging as $0\leq\vartheta\leq \pi$,
$0\leq\varphi\leq 2\pi$ and $0\leq\psi\leq {4\pi}$, we have the realization
\begin{eqnarray}
\sigma_1 & = &\sin\vartheta \sin\psi \, \mathrm{d}\varphi+\cos \psi \,
\mathrm{d}\vartheta \nonumber\\
\sigma_2 &=& \sin\vartheta\cos\psi\, \mathrm{d}\varphi-\sin\psi\, \mathrm{d}\vartheta\\
\sigma_3 &=& \cos\vartheta\, \mathrm{d}\varphi+\mathrm{d}\psi ~. \nonumber
\end{eqnarray}
These metrics are not isotropic in general.
The isometry group is enhanced to $SU(2) \times U(1)$ when any two $\gamma_i$'s coincide,
forming axisymmetric configurations, and it extends to $SU(2)\times SU(2)$ in the fully
isotropic case when all $\gamma_i$'s are equal.

The Ricci tensor $R_{ij}$, the Cotton tensor $C_{ij}$ and the fourth order tensor $K_{ij}$
given by equation \eqn{KK} are all diagonal in this base. Their non-vanishing components
take the form,
\begin{eqnarray}
R_{11}&=& \frac{1}{2\gamma_2 \gamma_3} \left[
\gamma_1^2 -(\gamma_2 - \gamma_3)^2 \right]\label{ric} ~, \label{R11} \\
C_{11}&=& -\frac{\gamma_1}{2( \gamma_1 \gamma_2 \gamma_3)^{\nicefrac{3}{2}}}
\left[\gamma^2_1\left(2\gamma_1-\gamma_2-\gamma_3\right)
-\left(\gamma_2+\gamma_3\right)\left(\gamma_2-\gamma_3\right)^2 \right] ~,\label{C11}
\label{cot}\\
K_{11}&=&  -
{\gamma_1 \over 32(\gamma_1 \gamma_2 \gamma_3)^2}
\left[21(5\gamma_1^4-3\gamma_2^4-3\gamma_3^4) - 2(\gamma_1^2\gamma_2^2+
\gamma_1^2\gamma_3^2-3 \gamma_2^2\gamma_3^2)-\right. \\
&&\left. 20\left(3\gamma_1^3(\gamma_2+\gamma_3)
-\gamma_2^3(\gamma_1+3\gamma_3)
-\gamma_3^3(\gamma_1+3\gamma_2)-\gamma_1\gamma_2\gamma_3(\gamma_1-\gamma_2-\gamma_3)
\right)\right] \nonumber
\end{eqnarray}
and there are similar expressions for the other components that follow by
cyclic permutation of the indices in all three tensors. Also, the Ricci scalar
curvature is given by
\begin{equation}
R = \frac{1}{2\gamma_1\gamma_2 \gamma_3} \left[
2\gamma_1\gamma_2 +2\gamma_2 \gamma_3+2\gamma_3 \gamma_1-\gamma_1^2 -\gamma_2^2-\gamma_3^2
\right]\label{Rsc} ~,
\end{equation}
whereas the trace of $K_{ij}$ is
\ba
K &=& {1\over32(\gamma_1\gamma_2\gamma_3)^2}\left[21(\gamma_1^4+\gamma_2^4+\gamma_3^4)
-2(\gamma_1^2\gamma_2^2+\gamma_1^2\gamma_3^2+\gamma_2^2\gamma_3^2) +\right.\\
&&\left. 20\left(\gamma_1\gamma_2\gamma_3(\gamma_1+\gamma_2+\gamma_3)-\gamma_1^3
(\gamma_2+\gamma_3) - \gamma_2^3(\gamma_1+\gamma_3) - \gamma_3^3(\gamma_1+\gamma_2)
\right)\right]\nonumber\,.
\ea

The expressions for the components of the Cotton tensor are obtained choosing
a particular orientation on $S^3$. For the opposite orientation these expressions
flip sign as they are odd under parity.

\subsection{Vacua of generalized massive gravity}

We are now in position to examine the reduced field equations and provide the
general solution of the resulting algebraic equations for generalized massive
gravity on $S^3$. The free parameters $\Lambda$, $\omega$ and $m^2$ are taken to
assume any real value at first. We will find three different classes of metrics,
in general: maximally isotropic with $SU(2) \times SU(2)$ isometry, axially
symmetric with $SU(2) \times U(1)$ isometry and totally anisotropic metrics with $SU(2)$
isometry. Restrictions on the range of free parameters will be placed in each
case separately so that the corresponding metrics have physical (Euclidean)
signature.

{\bf Isotropic solutions}:  Setting $\gamma_1 = \gamma_2 = \gamma_3 = \gamma$,
the field equations reduce to a single algebraic equation which has at most two
physically acceptable solutions
\be
\label{iso}
\gamma_\pm = {1 \over 8\Lambda} \left(1\pm \sqrt{1-{\Lambda \over m^2}} \right) ~.
\ee
Note that these maximally symmetric solutions are conformally flat and they
are independent of $\omega$, since the Cotton tensor vanishes.
For $m^2 =\Lambda$ there is a single solution with $\gamma = 1/8\Lambda$, which
only makes sense for $\Lambda >0$ (and hence $m^2 > 0$). For $\Lambda=0$ the only
physically acceptable solution is $\gamma_- = 1/16m^2$ when $m^2 > 0$. On the
other hand, for $m^2 > \Lambda > 0$, both solutions \eqn{iso} are real and positive,
thus leading to isotropic metrics with two different radii. Finally, for
$\Lambda > 0 > m^2$ or $m^2 > 0 > \Lambda$ only $\gamma_+$ and $\gamma_-$,
respectively, are physical solutions. In all other cases there are no
isotropic vacua in the theory. All these solutions are clearly isotropic solutions
of new massive gravity as well, since they are inert to the Chern-Simons
coupling $\omega$ by their symmetry.

{\bf Axially symmetric solutions}: Imposing axial symmetry sets two
coefficients of the metric equal, say $\gamma_1 = \gamma_2 \neq \gamma_3$, resulting to
partially anisotropic metrics on $S^3$ also known as Berger spheres or
bi-axially squashed spheres. By permuting
the coefficients $\gamma_i$ one can choose the axis of symmetry along any
principal direction of space, which can rotate into each other by $\mathbb{Z}_3$
symmetry. The corresponding metric takes the following form in terms of Euler angles,
\be
\mathrm{d}s^2= \gamma_1(\mathrm{d}\vartheta^2+\sin^2\vartheta \mathrm{d}\varphi^2) +
\gamma_3(\mathrm{d}\psi+\cos\vartheta \mathrm{d}\varphi)^2
\label{axialS}
\ee
and can be viewed as $S^1$ fibration over the base space $S^2$ with squashing
parameter $\gamma_3 / \gamma_1$.

The most efficient way to construct such solutions is by expressing the free
parameters $\omega$ and $\Lambda$ in terms of the metric coefficients and $m^2$;
of course, for any given solution, the parameters $\Lambda$ and $m^2$ are related
by the trace part of the field equations, $m^2 (R-6\Lambda) + K = 0$.
Then, for axially symmetric metrics, one finds that the most general solution is
determined by the special equations
\ba
\omega &=& {12 m^2\gamma_1\sqrt{\gamma_3}\over 4\gamma_1(1-2m^2\gamma_1)-21\gamma_3}\,,
\label{omegax}\\
\Lambda &=& {64m^2\gamma_1^3-40\gamma_1\gamma_3+21\gamma_3^2+16\gamma_1^2
(1-m^2\gamma_3)\over 192m^2\gamma_1^4}\label{Lax}\,,
\ea
which are valid for any non-zero value of $m^2$. It is possible to invert these
relations to express $\gamma_i=\gamma_i(m^2,\omega,\Lambda)$, but the resulting
expressions are quite lengthy and not very illuminating in general. For the
special cases of new massive gravity and topologically massive gravity these
expressions are much simpler and will be presented later.

There are at most three
axially symmetric solutions of the generalized massive gravity, depending on the
range of parameters, and, in particular, on the sign of $\omega$ that flips under parity.
Thus, for $m^2>\Lambda \ge 0$, there exist no axially symmetric solutions with $\omega>0$
and there are up to three distinct solutions with $\omega<0$, which, however, depend on the
particular value of $\omega$. Also, for $0<m^2<\Lambda$, the field equations admit one or two
axially symmetric solutions for positive or negative values of $\omega$, respectively.

{\bf Totally anisotropic solutions}: Configurations of this type are often called
tri-axially squashed spheres, and, as it turns out, they only exist for couplings satisfying
some very special relations. The first one is most conveniently described as
\be
{\omega \over m^2} =  {\left(2(\gamma_1\gamma_2+\gamma_1\gamma_3+\gamma_2\gamma_3)-
\gamma_1^2-\gamma_2^2-\gamma_3^2\right) \over 2 \left(\gamma_1^3+\gamma_2^3+
\gamma_3^3-(\gamma_1+\gamma_2)(\gamma_1+\gamma_3) (\gamma_2+\gamma_3)\right)}
\sqrt{\gamma_1\gamma_2\gamma_3}\,
\label{OManis}
\ee
with $\gamma_1 \neq \gamma_2 \neq \gamma_3$; the
cosmological constant $\Lambda$ is not arbitrary but it can be determined via
the trace part of the field equations $m^2 (R-6\Lambda) + K = 0$ in terms of the
corresponding $\gamma_i$.
Alternatively, using the expression of the Ricci scalar curvature \eqn{Rsc} and
introducing the following cubic combination of the metric coefficients,
\be
Y = \gamma_1^3+\gamma_2^3+\gamma_3^3-(\gamma_1+\gamma_2)(\gamma_1+\gamma_3)
(\gamma_2+\gamma_3)
\label{yy}\,,
\ee
we find that the totally anisotropic metrics satisfy the relation
\be
{\omega \over m^2} = (\gamma_1\gamma_2\gamma_3)^{3/2}{R\over Y} \,.
\label{mouriss}
\ee
This form is particularly useful for discussing the anisotropic solutions
of new massive gravity and topologically massive gravity as special cases.

It also turns out that such vacua satisfy the special curvature condition,
following from the equations,
\be
5R^2 = 16m^2 \Lambda\,
\label{con}
\ee
assuming non-vanishing (but finite) values of $\Lambda$ and $m^2$.
This relation may be valid for some axially symmetric solutions as well,
but not in general. It implies, in
particular, that $m^2 \Lambda > 0$ is a necessary condition for the existence of
totally anisotropic solutions; as for the parameter $\omega$, it can acquire positive
or negative values without spoiling the Euclidean signature of space. Using the
trace part of the field equations, we may eliminate $\Lambda$ and arrive at the
equivalent special relation $15 R^2 - 8m^2 R - 8K = 0$. Solving for $m^2$, we obtain
\ba
m^2 &=& {8(\gamma_1+\gamma_2+\gamma_3) \over \gamma_1^2+\gamma_2^2+\gamma_3^2 -2(\gamma_1\gamma_2+\gamma_1\gamma_3+
\gamma_2\gamma_3)} + \nonumber\\
&& {3(\gamma_1^2+\gamma_2^2+\gamma_3^2)+26(\gamma_1\gamma_2+\gamma_1\gamma_3+\gamma_2\gamma_3)
\over 8\gamma_1\gamma_2\gamma_3}~,
\label{NMGmu}
\ea
which together with condition \eqn{OManis} (or \eqn{mouriss}) provide the most efficient
way for the general description of such metrics.
The common points of these algebraic relations subsequently determine
$\gamma_i$ in terms of the free parameters of the theory, but the resulting expressions are
incredibly long and they are omitted from the presentation.

The general characteristics of all homogeneous solutions of generalized massive gravity
are summarized in the table below. All other cases do not materialize.

\vskip0.6cm
\begin{table}[h]
\centering
\begin{tabular}{|c|c|c|c|c|c|}
\hline
\textbf{Ricci}  & \textbf{Chern--Simons}  & \textbf{BHT mass}  & \textbf{totally}
& \textbf{axially}  & \multirow{2}{*}{\textbf{isotropic}}\\
\textbf{curvature} & \textbf{parameter} &  \textbf{parameter} & \textbf{anisotropic}
& \textbf{symmetric} &  \\
\hline
$R>0$ & $\omega>0$ & $m^2>0$ & no & $\gamma_1>\gamma_3$ & yes \\
\hline
$R>0$ & $\omega>0$ & $m^2<0$ & yes & yes & yes \\
\hline
$R>0$ & $\omega<0$ & $m^2>0$ & yes & yes & yes \\
\hline
$R>0$ & $\omega<0$ & $m^2<0$ & yes & yes & yes \\
\hline
$R<0$ & $\omega>0$ & $m^2<0$ & no & $\gamma_1<\gamma_3$ & no \\
\hline
$R<0$ & $\omega<0$ & $m^2>0$ & yes & $\gamma_1<\gamma_3$ & no\\
\hline
$R<0$ & $\omega<0$ & $m^2<0$ & no & $\gamma_1<\gamma_3$ & no \\
\hline
\end{tabular}
\vskip0.4cm
\caption{Geometric characteristics of vacua of generalized massive gravity.}
\end{table}

\subsection{Vacua of new massive gravity}

We now turn our attention to the corresponding solutions of new massive gravity,
which are obtained by taking the limit $|\omega|\to \infty$. There are homogeneous
solutions with all possible degrees of anisotropy, as in the general case.

{\bf Isotropic solutions}: The isotropic solutions are independent of $\omega$.
As such, they are identical to those of generalized massive gravity with
the same range of the parameters $m^2$ and $\Lambda$.

{\bf Axially symmetric solutions}: In this case, the equations \eqn{omegax} and \eqn{Lax}
simplify and yield at most two distinct axially symmetric metrics with coefficients
given explicitly by
\ba
\gamma_1 &=& \gamma_2 = {4m^2 \pm \sqrt{3m^2(5m^2+7\Lambda)} \over m^2
(21\Lambda -m^2)}\,,\\
\gamma_3 &=& {24m^2 (7\Lambda-11m^2) \pm 4 (21\Lambda-17m^2)\sqrt{3m^2(5m^2+7\Lambda)}
\over 21m^2 (21\Lambda -m^2)^2}\,.
\ea
For positive cosmological constant only the solution with the plus sign is physically
acceptable provided that $\Lambda > m^2 > 0$, whereas for $m^2<-7\Lambda/5$ there are two
axially symmetric solutions. Otherwise, there are no axially symmetric vacua
with $\Lambda > 0$. For negative cosmological constant only the solution with the
plus sign is physically acceptable provided that $21\Lambda< m^2 <0$, whereas for all
other values $m^2 <0$ there are two axially symmetric solutions.
Furthermore, there are no axially symmetric vacua when $m^2>0$ and $\Lambda<0$.

{\bf Totally anisotropic solutions}: The theory admits totally anisotropic solutions
provided that both $m^2$ and $\Lambda$ are negative. Such solutions are required to satisfy
the special condition $Y=0$,
\be
\gamma_1^3+\gamma_2^3+\gamma_3^3-(\gamma_1+\gamma_2)(\gamma_1+\gamma_3)(\gamma_2+\gamma_3)=0\,
\label{NMGcon}
\ee
that results from equation \eqn{mouriss} as $|\omega| \to \infty$.
The totally anisotropic solutions
have positive Ricci scalar curvature that is related to the parameters $\Lambda$ and $m^2$
by equation \eqn{con}, which is left intact by the limiting procedure. Alternatively, solving
for $m^2$, as in generalized massive gravity, the totally anisotropic vacua are determined by
the common points of \eqn{NMGcon} and \eqn{NMGmu} with unequal coefficients $\gamma_i$.

Thus, fixing $m^2 < 0$, we can describe all totally anisotropic vacua of new massive
gravity in parametric form. Solving for $\gamma_i (m^2, \Lambda)$ results into some
very complicated and lengthy expressions that are also omitted from the presentation.

The general characteristics of all homogeneous solutions of new massive gravity
are summarized in the next table:

\vskip0.6cm
\begin{table}[h]
\centering
\begin{tabular}{|c|c|c|c|c|}
\hline
\textbf{Ricci}  &  \textbf{BHT mass}  & \textbf{totally}  & \textbf{axially}
& \multirow{2}{*}{\textbf{isotropic}}\\
\textbf{curvature} &   \textbf{parameter} & \textbf{anisotropic}
& \textbf{symmetric} &  \\
\hline
$R>0$  & $m^2<0$ & yes & yes & yes  \\
\hline
$R>0$  & $m^2>0$ & no & $\gamma_1>\gamma_3$ & yes  \\
\hline
$R<0$  & $m^2<0$ & no & $\gamma_1<\gamma_3$ & no  \\
\hline
\end{tabular}
\vskip0.4cm
\caption{Geometric characteristics of vacua of new massive gravity.}
\end{table}

\subsection{Vacua of topologically massive gravity}

The homogeneous vacua of topologically massive gravity follow from the general
discussion by specializing the results to the limiting case $|m^2| \rightarrow \infty$.
As before, we have solutions with all possible degrees of anisotropy that are listed below.

{\bf Isotropic solutions}: The isotropic metrics are conformally flat, since their Cotton
tensor vanishes, and, therefore the only solution is
\be
\gamma = {1 \over 4 \Lambda} ~,
\ee
provided that $\Lambda > 0$, as in pure Einstein gravity. The positive cosmological
constant sets the scale for having a constant curvature (round) metric on $S^3$ with radius
$\sim 1/\sqrt{\Lambda}$.

{\bf Axially symmetric solutions}: In this case, the system of
equations \eqn{omegax} and \eqn{Lax} simplify to
\be
\omega = -{3\sqrt{\gamma_3} \over 2\gamma_1}\,, ~~~~~~
\Lambda = {4\gamma_1 -\gamma_3 \over 12\gamma_1^2}\,,
\ee
which, in turn, can be easily solved to yield the metric coefficients of
the axially symmetric metrics as
\be
\gamma_1 =  \gamma_2 = {9 \over \omega^2 + 27 \Lambda}\,  ,\quad \quad
\gamma_3 = {36 \omega^2 \over \left(\omega^2 + 27 \Lambda \right)^2} \,.
\label{TMGax}
\ee
Note that these solutions always exist provided that $\omega$ assumes negative
values (with the given choice of orientation made in section 3.1) without
other restriction. It can be readily seen from the expression for $\Lambda$
given above that axially symmetric metrics with $4\gamma_1 > \gamma_3$ correspond to
positive $\Lambda$ and, hence, to positive Ricci scalar curvature $R$, whereas
for $4\gamma_1 < \gamma_3$ the cosmological constant and the scalar Ricci
curvature are both negative. This is a well known property of the Berger spheres,
as one can flip the sign of $R$ by elongating the sphere beyond a critical value.
At the critical point $4 \gamma_1 = \gamma_3$ the curvature vanishes and so
does $\Lambda$ in the theory.

{\bf Totally anisotropic solutions}: By the same token, the totally anisotropic
solutions of topologically massive gravity can be easily obtained. Taking
$|m^2| \to\infty$, we find that these configurations have vanishing Ricci scalar
curvature, following from equation \eqn{mouriss}, in which case $\Lambda = 0$
and the special condition \eqn{con} is still satisfied in a limiting sense, as
$R=0$. In this limit, it can also be seen that $Y = 4\gamma_1\gamma_2\gamma_3$,
following from equation \eqn{yy} by taking into account the vanishing of the Ricci
scalar curvature of such metrics.
In addition, equation \eqn{OManis} yields the following expression for
the Chern--Simons coupling
\be
\omega = -{\gamma_1+\gamma_2+\gamma_3 \over \sqrt{\gamma_1\gamma_2\gamma_3}}\,,
\label{OManisTMG}
\ee
which is in fact required to be negative (with the given choice of orientation)
without further restriction.
Thus, the totally anisotropic vacua are described as common solutions of $R=0$ and
the condition \eqn{OManisTMG} demanding that $\gamma_1 \neq \gamma_2 \neq \gamma_3$.
We note that even in this limiting case it is not easy to express the metric
coefficients $\gamma_i$ in terms of $\omega$ in closed form. Note, however, that
this class of metrics has common element with the class of axially symmetric solutions
the Berger sphere with coefficients $\gamma_1 = \gamma_2 = \gamma_3 /4$.

The general characteristics of all homogeneous solutions of topologically massive gravity
are summarized in the table below:

\vskip0.6cm
\begin{table}[h]
\centering
\begin{tabular}{|c|c|c|c|c|}
\hline
\textbf{Ricci}  &  \textbf{Chern--Simons}  & \textbf{totally}  & \textbf{axially}
& \multirow{2}{*}{\textbf{isotropic}}\\
\textbf{curvature} &   \textbf{parameter} & \textbf{anisotropic}
& \textbf{symmetric} & \\
\hline
$R>0$  & $\omega>0$ & no & no & yes \\
\hline
$R>0$  & $\omega<0$ & no & yes & yes \\
\hline
$R=0$  & $\omega<0$ & yes & $\gamma_1<\gamma_3$ & no \\
\hline
$R<0$  & $\omega<0$ & no & $\gamma_1<\gamma_3$ & no \\
\hline
\end{tabular}
\vskip0.4cm
\caption{Geometric characteristics of vacua of topologically massive gravity.}
\end{table}

The classification of all homogeneous vacua of topologically massive gravity
was carried out in the literature long time ago, including the totally anisotropic
solutions, \cite{vuorio, Nutku:1989qi,Ortiz:1989vc} (but see also \cite{Chow:2009km}
for an overview, as well as the more recent work \cite{bakas}). This justifies the
parametrization used earlier for the
presentation of the classification scheme of all
homogeneous solutions of generalized massive gravity (see, in particular, equation
\eqn{OManis}) and it is gratifying to see how these special results are reproduced
from our general construction.

\subsection{Other special limiting cases}

Concluding this section, we discuss three special limiting cases that arise in the
space of couplings. The homogeneous solutions we obtain in these cases are rather simple
and they form the basis for the more general vacua that arise by competition of the
individual terms in the general theory.

First, by taking the limit $\kappa_{\rm w} \rightarrow 0$, we obtain pure Einstein
gravity that exhibits a fully isotropic solution for $\Lambda > 0$, so that $R=6\Lambda$,
and there are no other homogeneous vacua. Next, by taking the limit $\omega \rightarrow 0$,
we obtain the pure Cotton theory of conformal gravity, \cite{horne}, which
exhibits a fully isotropic solution and a degenerate
axially symmetric vacuum with $\gamma_1 = \gamma_2 = \infty$ and $\gamma_3 = 0$,
which is nevertheless regular provided that the volume of space ($\sim \sqrt{\gamma_1
\gamma_2 \gamma_3}$) is held finite. The latter metric corresponds to a fully squashed
configuration along one of the principal directions of $S^3$ and it is unique
up to permutations of the axes. Finally, pure fourth order gravity follows from the
general theory in the limit $m \rightarrow 0$, \cite{standes}. In this case, by first
considering the traceless part of the equations of motion, we find that there is a
fully isotropic solution as well as two different axially symmetric solutions, which
are unique up to permutations of the axes. One of them is the degenerate (but regular)
metric on the fully squashed $S^3$, as in the pure Cotton theory, and the other is
a non-degenerate Berger sphere with $\gamma_1 = \gamma_2$ and $\gamma_3 = 4 \gamma_1/21$.
However, non of these metrics satisfies the trace part of the classical equations of motion,
$K=0$, and, therefore, pure fourth order gravity has no regular homogeneous vacua; this is
also consistent with the absence of a length scale in the model that can stabilize the
vacua, if any.

Turning on all parameters in generalized massive gravity allows for more complex
situations that can balance the effect of different terms and produce the web of
the homogeneous vacua we have described above. In all cases, it is convenient to
first solve the traceless part of the classical equations of motion and then examine
the constraints imposed on the vacua from the trace of the equations to obtain
physically acceptable solutions with different characteristics. The traceless
part of the equations of motion will also be used in the next section to provide
an algebraic classification of the homogeneous metrics on $AdS_3$ following by
analytic continuation.

\section{Homogeneous solutions on AdS$_3$}
\setcounter{equation}{0}

Analytic continuation of the squashed spheres yields homogeneous solutions
of massive gravity on $AdS_3$ with Lorentzian signature. $S^3$ is an $S^1$
fibration over $S^2$ and, therefore, there are two inequivalent ways to
obtain squashed metrics on $AdS_3$ depending on the choice of time-like direction,
$\tau$.
One is associated to time-like squashed metrics by viewing $AdS_3$ as time-like
fibration over the hyperbolic plane $H_2$ and the other to space-like squashed
metrics by viewing $AdS_3$ as space-like fibration over $AdS_2$ space.
We will consider both possibilities below and indicate how the classification
of homogeneous vacua on $S^3$ carry to homogeneous metrics on $AdS_3$ with the
appropriate choice of coupling constants in the theory.

First, we consider the following analytic continuation and define coordinates
$\tau$, $\rho$ and $z$ as
\be
\psi = \tau ~, ~~~~~ \vartheta = {\pi \over 2} - i \rho ~, ~~~~~ \varphi = -i z ~.
\ee
Then, the metric on $AdS_3$ with positive definite coefficients $\gamma_1$,
$\gamma_2$, $\gamma_3$ takes the general form
\ba
ds^2 & = & \gamma_1 \left({\rm cos} \tau ~ d\rho + {\rm sin}\tau ~ {\rm cosh} \rho ~
dz \right)^2 + \gamma_2 \left({\rm sin} \tau ~ d\rho - {\rm cos}\tau ~ {\rm cosh} \rho
~ dz \right)^2 - \nonumber\\
& & \gamma_3 \left(d\tau + {\rm sinh} \rho ~ dz \right)^2
\ea
after flipping the overall sign of the metric to have signature $-++$.
For axially symmetric configurations with $\gamma_1 = \gamma_2$ it specializes to
\be
ds^2 = \gamma_1 \left(d\rho^2 + {\rm cosh}^2 \rho ~ dz^2 \right) -
\gamma_3 \left(d\tau + {\rm sinh} \rho ~ dz \right)^2 ~.
\ee
This corresponds to the case of time-like squashing, where the base space is $H_2$
with metric $d\rho^2 + {\rm cosh}^2 \rho ~ dz^2$. If we had considered, instead, the
analytic continuation $\vartheta = i \rho$, $\psi = \tau$ and $\varphi = z$,
the hyperbolic trigonometric functions ${\rm sinh} \rho$ and ${\rm cosh} \rho$
would have been exchanged, resulting to time-like squashing of $AdS_3$ over the
base space $H_2$ with metric $d\rho^2 + {\rm sinh}^2 \rho ~ dz^2$. The two choices
are clearly related to each other as they correspond to different coordinate patches
on $H_2$ (often called hyperbolic and elliptic solutions, respectively). Here, we
choose to work with the first one as it provides global coordinates in space.

Next, we consider another analytic continuation by defining coordinates
$\tau$, $\rho$ and $z$ as follows,
\be
\varphi = \tau ~, ~~~~~ \vartheta = {\pi \over 2} - i \rho ~, ~~~~~ \psi = i z ~,
\ee
which provide a different choice of time-like direction as the role
of $\tau$ and $z$ coordinates is exchanged. Then, the general
homogeneous metric on $AdS_3$ takes the form
\ba
ds^2 = & = & \gamma_1 \left({\rm cosh} z ~ d\rho - {\rm sinh}z ~ {\rm cosh} \rho ~
d\tau\right)^2 - \gamma_2 \left({\rm sinh} z ~ d\rho - {\rm cosh}z ~ {\rm cosh} \rho
~ d\tau\right)^2 + \nonumber\\
& & \gamma_3 \left(dz + {\rm sinh} \rho ~ d\tau \right)^2 ,
\ea
which specializes for $\gamma_1 = \gamma_2$ to the bi-axially squashed metrics
\be
ds^2 = \gamma_1 \left(d\rho^2 - {\rm cosh}^2 \rho ~ d\tau^2\right) +
\gamma_3 \left(dz + {\rm sinh} \rho ~ d\tau \right)^2 ~.
\ee
As before, we also flip the overall sign of the metric to have signature $-++$.
These correspond to metrics on $AdS_3$ with space-like squashing, since the
base space is $AdS_2$ with metric $d\rho^2 - {\rm cosh}^2 \rho ~ d\tau^2$.
We also note for completeness that another choice of coordinate patch on base
space with metric $d\rho^2 - {\rm sinh}^2 \rho ~ d\tau^2$ would have resulted
from the analytic continuation $\vartheta = i \rho$, $\varphi = \tau$ and
$\psi = z$.

In all cases we obtain $AdS_3$ vacua with coefficients and parameters given as
before for all different types of squashing in the generalized
theory of three-dimensional massive gravity and its simpler variants (new and
topologically massive gravity), thus making the repetition of equations obsolete.
The only difference that should be taken into account, as compared to the
corresponding expressions for the homogeneous vacua found in section 3, is that the
cosmological constant $\Lambda$ should be replaced by $-\Lambda$, since we have
chosen the signature $-++$ on $AdS_3$ using a sign flip of the metric after
analytic continuation; if we had
chosen to work with signature $+--$ this would not be needed. Likewise,
the other parameters $\omega$ and $m^2$ also flip sign and they should be
replaced by $-\omega$ and $-m^2$. Note, however, that the time-like and space-like
squashed metrics are mutually related by exchanging the role of $\varphi$ and $\psi$
coordinates on $S^3$ (prior to analytic continuation), which in turn imply a
change of orientation. Thus, if $\omega$ is replaced by $-\omega$ in the case
of time-like squashed metrics on $AdS_3$, as explained above, $\omega$ will not
flip sign in the space-like squashed metrics.
With these explanations in mind we obtain complete classification of all
homogeneous metrics on $AdS_3$ with $SU(1, 1)$ isometry group. Their existence
and tabulation as time-like and space-like squashed vacua follows easily from
the corresponding tables found in section 3 with the appropriate range of
parameters. For the simpler case of topologically massive gravity, which has
been studied for a long time, the results are in agreement with those reported
in earlier works on the subject, \cite{Nutku:1989qi, Ortiz:1989vc} (but see also
\cite{Chow:2009km} for an overview and many more references to the literature).

Concluding this section, we comment on the algebraic characterization of the
space-time metrics on $AdS_3$ based on the Petrov and Segre classification
(see, for instance, \cite{ExactSol} for the general scheme as it was initially developed
in four space-time dimensions). In three dimensions, the Petrov classification refers
to the Cotton tensor ${C^i}_j$, \cite{Barrow, Hall1, Torres, Garcia, Sousa}, and the
Segre classification refers to the traceless Ricci tensor ${S^i}_j =
{R^i}_j - {\delta^i}_j ~R/3$, \cite{Hall2, Hall1, Torres, Sousa}. In either case one
views these second rank tensors as linear maps between 3-vectors which are classified
according to the number of distinct eigenvalues and the space-time character of
their eigenvectors; we refer to the literature for the details and notation used for
the different classes of space-times.

Topologically massive gravity is rather special in this context, because the Petrov and
Segre classifications coincide by the traceless part of the classical equations of motion
\eqn{TMGe2}, although the notation used in the literature depends on the particular scheme.
It is rather instructive to briefly summarize the results of the algebraic characterization
of all homogeneous vacua of topologically massive gravity, following \cite{Chow:2009km}.
In this case, determining the eigenvalues of ${S^i}_j$ and their multiplicities is
equivalent to finding the scalar invariants
\be
I = {S^i}_j {S^j}_i = {\rm tr}(S^2) ~, ~~~~~~
J = {S^i}_j {S^j}_k {S^k}_i = {\rm tr}(S^3) ~.
\ee
Bi-axially squashed $AdS_3$ metrics are of Petrov type $D$ and one often distinguishes
between the time-like and space-like squashed metrics using the notation $D_{\rm t}$ and
$D_{\rm s}$ respectively. These solutions are denoted by $[(11),1]$ and $[1(1,1)]$,
respectively, in the Segre classification scheme and satisfy the relation
$I^3 = 6J^2 \neq 0$. Isotropic solutions are of Petrov type $O$ and of Segre type $[(11, 1)]$
satisfying the special relation $I=J=0$. Finally, totally anisotropic metrics are of Petrov type
$I_\mathbb{R}$ and of Segre type $[11, 1]$ satisfying the relation $I^3 > 6J^2$. Of course,
it is also possible to have other solutions of more general algebraic type but they fall
outside the class of homogeneous metrics and we are not going to discuss these here.

For the generalized massive gravity, and its limiting theory of new massive gravity, the
three-dimensional analogues of the Petrov and Segre classifications are distinct because
$C_{ij}$ is no longer proportional to $S_{ij}$. Still one can classify their homogeneous
solutions into algebraic types, which turn out to be identical to those appearing in
topologically massive gravity. This can be explicitly checked case by case for $AdS_3$ vacua
with all possible degrees of anisotropy and verify that they are of Petrov type $O$, $D$
($D_{\rm t}$ or $D_{\rm s}$) and $I_\mathbb{R}$. Likewise, one can characterize these
vacua by their Segre type and find exactly the same classes in the notation used above.

\section{Applications to $z=4$ Ho\v{r}ava--Lifshitz gravity}
\setcounter{equation}{0}

We will discuss some applications of our results to Ho\v{r}ava--Lifshitz gravity
in $3+1$ dimensions. This is a non-relativistic theory of gravitation that has been
proposed as ultra-violet completion of Einstein's theory, \cite{horava}, but it also
serves as toy model for transitions among vacua of three-dimensional gravity in the
spirit of Onsager--Machlup theory for non-equilibrium processes, \cite{onsa}.

Space-time is assumed to be $M_4 = \mathbb{R} \times \Sigma_3$ and the theory is defined
using the ADM (Arnowitt--Deser--Misner) decomposition of the metric
\begin{equation}
\mathrm{d}s^2 = -N^2 \mathrm{d}t^2 +g_{ij}\left(\mathrm{d}x^i +N^i \mathrm{d}t\right)
\left(\mathrm{d}x^j +N^j \mathrm{d}t\right) .
\end{equation}
The metric on the spatial slices $\Sigma_3$ is $g_{ij}$, whereas $N$ and $N^i$ are the
lapse and shift functions, respectively, which depend on all space-time coordinates,
in general\footnote{We use Latin indices $i, j, \cdots$ to indicate that
$\Sigma_3$ is always Riemannian here. Before we used Greek indices $\mu, \nu, \cdots$
to allow for both Riemannian and pseudo-Riemannian metrics in the discussion of
three-dimensional gravitational theories.}.
The infinite dimensional space of all three-dimensional Riemannian metrics
$g_{ij}$ is called superspace and it is endowed with a metric
\begin{equation}
\label{dwm}
G^{ijk\ell}=\frac{1}{2}\left(g^{ik}g^{j\ell}+g^{i\ell}g^{jk}\right)-\lambda
g^{ij}g^{k\ell}
\end{equation}
that generalizes the standard DeWitt metric using an arbitrary parameter $\lambda$
(other than $1$). The inverse metric in superspace is
\begin{equation}
\mathcal{G}_{ijk\ell}=\frac{1}{2}\left(g_{ik}g_{j\ell}+g_{i\ell}g_{jk}\right)-
\frac{\lambda}{3\lambda-1} g_{ij}g_{k\ell}
\end{equation}
so that
\begin{equation}
G^{ijk\ell}\mathcal{G}_{k\ell mn}= \frac{1}{2}(\delta^i_m \delta^j_n +\delta^i_n
\delta^j_m ) ~.
\end{equation}

The action of Ho\v{r}ava--Lifshitz gravity in $3+1$ dimensions is written as a sum of
kinetic and potential terms. Assuming {\em detailed balance}, which is important for
our discussion, the action takes the form, \cite{horava},
\begin{equation}
S_{\rm HL}  =  {2 \over \kappa^2} \int dt d^3x \sqrt{g}N K_{ij} G^{ijk\ell} K_{k\ell} -
{\kappa^2 \over 2} \int dt d^3x  \sqrt{g}N E^{ij} \mathcal{G}_{ijk\ell} E^{k\ell} ~,
\end{equation}
where $K_{ij}$ is the second fundamental form measuring the extrinsic curvature
of the spatial slices $\Sigma$ at constant $t$ (not to be confused with the fourth
order tensor $K_{ij}$ of new massive gravity),
\begin{equation}
K_{ij}=\frac{1}{2N}\left(\partial _t g_{ij}-\nabla_iN_j-\nabla_jN_i\right)
\end{equation}
and
\begin{equation}
\label{eom}
E^{ij}=-\frac{1}{2 \sqrt{g}}\frac{\delta W[g]}{\delta g_{ij}} ~.
\end{equation}
The four-dimensional gravitational coupling is $\kappa$.
The kinetic term contains two time derivatives of the metric $g_{ij}$, and, as such,
it is identical to general relativity in canonical form (though $\lambda$ is taken
arbitrary here). The potential term is different, however, as it is derived
from a superpotential functional $W$ that is chosen appropriately to render
the theory power-counting renormalizable.

In the following, we choose $W[g]$ to be the action functional of Euclidean
three-dimensional massive gravity, setting, in general, $W = S_{\rm GMG}$. Then,
the theory has anisotropy scaling parameter is $z=4$, since the highest order term
in the potential of $S_{\rm HL}$ is $K_{ij} K^{ij}$ followed by $C_{ij} C^{ij}$
and $R_{ij} R^{ij}$ as well as other subleading cross terms, \cite{cai}.
We will restrict attention to the so called projectable case of Ho\v{r}ava-Lifshitz
gravity, meaning that the lapse function $N$ associated with the freedom of time
reparametrization is restricted to be a function of $t$, whereas the shift functions
$N_i$ associated with diffeomorphisms of $\Sigma_3$ can depend on all space-time
coordinates. In view of the applications that will be discussed next, we choose
\begin{equation}
N(t) = 1 ~, ~~~~~~ N^i (t, x) = 0~,
\end{equation}
without great loss of generality.

It is clear that the vacua of three-dimensional massive gravity provide static (i.e.,
$t$-independent) solutions of Ho\v{r}ava--Lifshitz gravity, which is one of the
applications. More importantly, these vacua can also be used to support instanton
solutions that interpolate smoothly between different critical points of $W[g]$, and,
hence, of the potential functional of the four-dimensional action $S_{\rm HL}$.
Although the description of instanton solutions will be quite general here, following
earlier work on the subject, \cite{bakas}, specialization to homogeneous vacua of
generalized massive gravity on $\Sigma_3 \simeq S^3$ leads to a classification
scheme for all $SU(2)$ gravitational instantons of Ho\v{r}ava--Lifshitz theory with
anisotropy scaling parameter $z=4$. It also puts the results of section 3 in a
wider context and makes them the basis for future developments. The rest of this section
outlines this construction, but more details will be presented elsewhere, \cite{sourdpap}.

Let us now consider the Euclidean action of Ho\v{r}ava--Lifshitz theory
which is obtained by analytic continuation in time. Furthermore, for technical reasons
that will become apparent in a moment, we restrict the parameter $\lambda$ of the
superspace metric in the range
\begin{equation}
\lambda < 1/3
\end{equation}
so that $G^{ijk \ell}$ is positive definite. Also, $\Sigma_3$ is assumed to be compact
with no boundary, as in the typical case $\Sigma_3 \simeq S^3$ we are considering here.
Then, the Euclidean action can be manipulated by standard elementary methods as follows,
\cite{bakas},
\begin{eqnarray}
S_{\rm HL}^{\rm Eucl} & = &
{2 \over \kappa^2} \int dt d^3x \sqrt{g} K_{ij} G^{ijk\ell} K_{k\ell} +
{\kappa^2 \over 2} \int dt d^3x \sqrt{g} E^{ij} \mathcal{G}_{ijk\ell} E^{k\ell}
\nonumber\\
& = & \frac{2}{\kappa^2}\int \mathrm{d}t\,\mathrm{d}^3x\,\sqrt{g}
\left(K_{ij}\pm \frac{\kappa^2}{2}\mathcal{G}_{ijmn} E^{mn}\right){G}^{ijk\ell}
\left(K_{k\ell}\pm \frac{\kappa^2}{2}\mathcal{G}_{k\ell rs} E^{rs}\right) \nonumber\\
& & \mp 2 \int dt d^3x \sqrt{g} K_{ij} E^{ij} ~,
\end{eqnarray}
taking into proper account all boundary terms. Thus, for positive definite
superspace metric, the Euclidean action appears to be bounded from below by
\begin{equation}
S_{\rm HL}^{\rm Eucl} \ge \mp 2 \int dt d^3x \sqrt{g} K_{ij} E^{ij} = \mp \int dt d^3x
\sqrt{g} E^{ij} \partial_t g_{ij} = \pm {1 \over 2} \int dt {dW \over dt} ~.
\end{equation}
Extrema of the action are provided by configurations satisfying the following
special equations that are first order in time,
\begin{equation}
\label{spezial}
K_{ij} \equiv \frac{1}{2} \partial _t g_{ij} = \mp  \frac{\kappa^2}{2}
\mathcal{G}_{ijmn} E^{mn} ~,
\end{equation}
which are the defining equations of instantons.

As the spatial slices evolve in Euclidean time following \eqn{spezial}, the
superpotential functional $W$ changes monotonically. This is easily seen by
considering
\begin{equation}
\label{bpsb}
{dW \over dt} = -2 \int d^3x \sqrt{g} E^{ij} \partial_t g_{ij} = \pm 2 \kappa^2
\int d^3x \sqrt{g} E^{ij} \mathcal{G}_{ijk\ell} E^{k\ell} ~,
\end{equation}
which is the integral of a quadratic quantity when $\lambda < 1/3$, and, therefore,
it increases or decreases monotonically depending on the overall sign. Using this
observation and by taking the time integral of equation \eqn{bpsb}, it turns out
that the lower bound of the Euclidean action $S_{\rm HL}$ is always positive and
it is saturated by the special configurations \eqn{spezial}. Then, the instanton
action is
\begin{equation}
S_{\rm HL}^{\rm Eucl} = {1 \over 2} |\Delta W| ~,
\end{equation}
where $\Delta W$ denotes the difference of the corresponding values of $W$ at the
two end points of the time interval that supports such solutions.
Instantons and anti-instantons are associated to the two different sign options,
and, therefore, they are mutually related by reversing the arrow of time.

As in ordinary instanton physics, it is also appropriate here to consider solutions
with finite Euclidean action only. This is possible provided that there are solutions of
equation \eqn{spezial} that extrapolate smoothly between degenerate minima of the
potential for, otherwise, $W$ may become infinite. This restriction is also imposed
by the space-time interpretation of the solutions of Euclidean Ho\v{r}ava--Lifshitz
gravity in order to obtain complete spaces with non-singular metrics. Thus, instantons
are naturally associated to eternal solutions of certain higher order geometric
flow equations, which are gradient flows of $W$ according to equation \eqn{spezial}.
In particular, setting $W=S_{\rm GMG}$, we obtain the following evolution equations,
\begin{eqnarray}
\label{fullflow}
\partial_t g_{ij} & = & - \frac{\kappa^2}{2}\left(
R_{ij} - {2\lambda - 1 \over 2(3\lambda -1)} R g_{ij} - \frac{\Lambda}{3\lambda -1}
g_{ij} \right)- \frac{\kappa^2 }{2\omega} C_{ij} \nonumber\\
& & + \frac{\kappa^2}{4 m^2} \left(K_{ij} - {\lambda \over 3\lambda -1} K \right)~,
\end{eqnarray}
choosing for definiteness one of the two sign options. The instanton solutions correspond
to trajectories that connect continuously any two fixed points of the flow, without
encountering singularities, as $-\infty < t < +\infty$. They are solely selected by their
boundary conditions, having spatial slices with zero extrinsic curvature (equal to the
normal derivative $\partial_t g_{ij}$) at the two end-points of their Euclidean
life-time\footnote{Despite appearances, the fixed points of the flow equation
\eqn{fullflow} are independent of $\lambda$ and coincide with the vacua of generalized
massive gravity. The parameter $\lambda$ only affects the form of the flow lines that
define the instantons.}.
All other flow lines of the geometric evolution equation \eqn{spezial} do not qualify
as instantons, and, in general, they become extinct (typically in finite time) by
encountering singularities, thus leading to infinite action $S_{\rm HL}^{\rm Eucl}$;
they are discarded from our general construction.

The explicit construction and classification of all instanton solutions of $z=4$
Ho\v{r}ava--Lifshitz theory relies heavily on two open problems. The first is the
classification of all vacua of generalized massive gravity, which serve as end-points
of the interpolating instanton metrics. The second is the general behavior of higher
order curvature flows, as \eqn{fullflow}, and the possible occurrence of singularities
that may inflict the trajectories. The standard methods that are available for
studying second order equations are not applicable any more and even the short
time existence of solutions is now questionable, in general. Focusing on homogeneous vacua
offers a mini-superspace model to study these problems and obtain concrete results.
We have obtained complete classification of all fixed points as classical
solutions on $S^3$ with $SU(2)$ symmetry and at the same time the flow equations
reduce consistently to a closed system of ordinary differential equations for the
metric coefficients $\gamma_i$ as functions of time. Then, eternal solutions of these
equations are in one-to-one correspondence with the $SU(2)$ gravitational instanton
solutions of $z=4$ Ho\v{r}ava--Lifshitz theory. Explicit constructions are possible
by extending previous results \cite{bakas} to fourth order flows, but the details
are more complicated and they will be presented in a separate paper.

The instanton solutions of Ho\v{r}ava--Lifshitz gravity can also be used to describe
off-shell transitions among the many different vacua that populate the landscape of
massive gravity models. This alternative interpretation is in the spirit of
Onsager-Machlup theory for non-equilibrium processes in thermodynamics, \cite{onsa}.
In this general context, $W$ is the entropy function that changes monotonically in time
and it is proportional to the logarithm of the probability of a given fluctuation.
The gradient of $W$ is the thermodynamic force measuring the tendency of a system to seek
equilibrium. Linearization of the flow equations around the fixed points describe
small fluctuations away from equilibrium states, whereas the instanton solutions
incorporate non-linear effects for large transitions between different states of
the system. It will be interesting to strengthen the analogy between non-equilibrium
processes and geometric flow equations by focusing, in particular, to three-dimensional
massive gravity models as working example and explore its higher dimensional origin by
embedding the theory in string or M-theory framework. A renormalization group approach
to the instanton solutions might also emerge from this study.

\section{Conclusions}
\setcounter{equation}{0}

We have classified all homogeneous vacua of (generalized) new massive gravity in
three dimensions using the Bianchi IX ansatz for Riemannian metrics on $S^3$.
We have also obtained the corresponding $AdS_3$ metrics by analytic continuation
and characterized them algebraically using the Petrov and Segre schemes. Our
results provide generalization of the homogeneous vacua of topologically massive
gravity in the presence of a new quadratic curvature term in the action based on
the recent proposal \cite{BHT}. In all cases we found that homogeneous metrics with
different degrees of anisotropy can be realized as vacua in certain regions of
the parameter space of couplings. The most exotic case is provided by the totally
anisotropic (i.e., tri-axially squashed) metrics, which have special Ricci scalar
curvature. Although the explicit form of the metric coefficients are rather
cumbersome to present, in general, as functions of the couplings, the action
takes particularly simple form, as can be found (but not shown here).
These critical values of the action can be used
to compute the instanton action of interpolating configurations among the
different vacua and associate them to a probability measure by advancing further
the connections with higher order geometric flows.

It should be emphasized that these homogeneous solutions coexist in certain
regions of the parameter space of couplings, as summarized in Tables 1, 2 and 3.
Thus, fixing the couplings $\omega$, $m^2$ and $\Lambda$ one may have isotropic,
axially symmetric and totally anisotropic configurations as distinct classical
solutions of three-dimensional gravity.
In other regions of the parameter space only some of these vacua can coexist.
They all provide the landscape of homogeneous vacua in mini-superspace, and, as
such, they are not continuously connected to each other. Some of these vacua can
coalesce by varying the couplings, in which case
their defining relations coincide for special values of $\omega$ and $m^2$, as can
be readily seen from the equations. These remarks apply to generalized new
massive gravity by extending previously known results for topologically massive
gravity. Finally, Ho\v{r}ava--Lifshitz gravity was used as toy model to study
off-shell transitions among these vacua. In this context, we were not concerned with
the shortcomings and problems of such alternative theories of gravitation (see for
instance \cite{blaso} for a recent overview and references therein), but we certainly
have to face them in detail when applying our results to Euclidean gravity. We hope
to return to these problems elsewhere.

In future work, it will be interesting to consider other classes of solutions
of generalized massive gravity and explore more regions in the landscape of
vacua. Although this is a rather intricate problem, it is the simplest to
address in the context of gravitational theories with propagating degrees of
freedom. Another important question is the possibility to embed such
three-dimensional theories in string or M-theory and use them to investigate
the structure of the corresponding space-time configurations in higher
dimensions. Although some partial results exist in this direction, in particular
for topologically massive gravity, \cite{luu}, the general framework is still
lacking. We hope to be able to report on this and related issues elsewhere.

\end{document}